\documentclass{article}

\usepackage{arxiv}
\usepackage{amsmath}
\usepackage[utf8]{inputenc} % allow utf-8 input
\usepackage[T1]{fontenc}    % use 8-bit T1 fonts
\usepackage{cite} % purely to disable citation links
\usepackage[hyperfootnotes=false,colorlinks=true,linkcolor=black,anchorcolor=black,citecolor=black,filecolor=black,menucolor=black,runcolor=black,urlcolor=black,breaklinks=true]{hyperref}       % hyperlinks
\usepackage{xurl}
\usepackage{url}            % simple URL typesetting
\usepackage{booktabs}       % professional-quality tables
\usepackage{amsfonts}       % blackboard math symbols
\usepackage{nicefrac}       % compact symbols for 1/2, etc.
\usepackage{microtype}      % microtypography
\usepackage{amssymb}
\usepackage{float}
\usepackage{listings}
\newsavebox{\LstBox}
\usepackage{graphicx}
\usepackage{gensymb} % for degree symbol
\usepackage{caption}
\usepackage[round]{natbib}
\usepackage{doi}
\usepackage{enumitem}
\usepackage{framed}
\usepackage[table]{xcolor}
\usepackage{colortbl}
\usepackage{courier}
\definecolor{highestcolor}{RGB}{153, 255, 153} % Dark green for highest values
\definecolor{highcolor}{RGB}{204, 255, 204}    % Light green for high values
\definecolor{midcolor}{RGB}{255, 255, 204}     % Light yellow for mid values
\definecolor{lowcolor}{RGB}{255, 229, 204}     % Light orange for low values
\definecolor{lowestcolor}{RGB}{255, 204, 204}  % Light red for lowest values

\definecolor{lightgray}{rgb}{0.95, 0.95, 0.95}
\definecolor{darkgray}{rgb}{0.4, 0.4, 0.4}

\colorlet{shadecolor}{lightgray}

\usepackage[OT2,OT1]{fontenc}
\newcommand\cyr
{
\renewcommand\rmdefault{wncyr}
\renewcommand\sfdefault{wncyss}
\renewcommand\encodingdefault{OT2}
\normalfont
\selectfont
}
\DeclareTextFontCommand{\textcyr}{\cyr}

\setlist[itemize]{leftmargin=*}
\defcitealias{houseequifax}{House Equifax Report}
\defcitealias{criminal}{United States v. Park Jin Hyok, Complaint, Case No. MJ18-1479 (C.D. Cal. 2018)}
\defcitealias{ai2027}{AI 2027}

\title{Uplifted Attackers, Human Defenders: The Cyber Offense-Defense Balance for Trailing-Edge Organizations}
\date{}
\author{Benjamin~Murphy\thanks{Harvard Law School}\quad Twm~Stone\thanks{Independent}}

\hypersetup{
pdftitle={Uplifted Attackers, Human Defenders: The Cyber Offense-Defense Balance for Trailing-Edge Organizations},
pdfsubject={cs.CR},
pdfauthor={Benjamin~Murphy, Twm~Stone},
colorlinks=true,
urlcolor=blue
}

\begin{document}
  
\maketitle
\vspace{-30pt}
\setcounter{footnote}{0}
\begin{abstract}
Advances in artificial intelligence are widely understood to have implications for cybersecurity. Articles have emphasized the effect of AI on the cyber offense-defense balance, and credible commentators can be found arguing either that cyber will privilege attackers or defenders. For defenders, arguments are often made that AI will enable solutions like formal verification of all software—and for some well-equipped companies, this may be true. This conversation, however, does not match the reality for most companies. ``Trailing-edge organizations,'' as we term them, rely heavily on legacy software, poorly staff security roles, and  struggle to implement best practices like rapid deployment of security patches. These decisions may be the result of corporate inertia, but may also be the result of a seemingly-rational calculation that attackers may not bother targeting a firm due to lack of economic incentives, and as a result, underinvestment in defense will not be punished.
\vspace{0.2\baselineskip}

This approach to security may have been sufficient prior to the development of AI systems, but it is unlikely to remain viable in the near future. We argue that continuing improvements in AI's capabilities poses additional risks on two fronts: First, increased usage of AI will alter the economics of the marginal cyberattack and expose these trailing-edge organizations to more attackers, more frequently. Second, AI's advances will enable attackers to develop exploits and launch attacks earlier than they can today—meaning that it is insufficient for these companies to attain parity with \textit{today's} leading defenders, but must instead aim for faster remediation timelines and more resilient software.
\vspace{0.2\baselineskip}

Trailing-edge organizations exist in a grim reality. Their minimal investment in cybersecurity has been premised on an assumption that attackers are insufficiently incentivized to target them. AI's effects on the economics and technical capacity of cyberattacks will expose these organizations to substantially heightened risk. This may spur additional investment in defense, but likely only after these organizations are subject to substantial damages. Our analysis points to a substantial degree of exposure across the economy, with only limited mitigating factors arising from AI's improvements to cyberdefense. The situation today portends a dramatically increased number of attacks in the near future, a reality which has not been captured by the existing discussion of AI-enabled cyber risks. Moving forward, we offer a range of solutions for both individual organizations and governments designed to improve the defensive posture of firms which lag behind their peers today.

\vspace{0.15cm}

\end{abstract}

\newpage
\section{Introduction}
\subsection{AI's Impact on Cyber}
The rate of innovation in artificial intelligence (AI) has produced a steady stream of warnings from the policy and academic communities. From advanced knowledge useful for bioweapons development to highly effective emotional manipulation, newly-developed models regularly demonstrate skills that frequently can be harnessed both for positive, economically productive ends, as well as for harmful ones.\footnote{\textit{See} International AI Safety Report \citep{ISRSAA2025}.} These attributes are frequently referred to as ``dual-use'' insofar as they can be used for either constructive or destructive purposes.\footnote{\citet{forbesdual}.} AI's dual-use nature is nowhere more evident than in cyber, where many authors have warned about a coming wave of cyberattacks, while others have correspondingly demonstrated how models can assist with cyberdefense by enabling faster detection of software vulnerabilities and automating the process of developing and deploying patches.\footnote{\textit{See, e.g.}, \citet{lohnimpact}, \citet{tang2024implications}.}
\vspace{0.2\baselineskip}

Cyber is a particularly fertile ground for AI-driven disruption for three reasons. First, there are independent economic rationales for improving AI's coding capabilities.\footnote{\textit{See, e.g.}, \citet{cursorvaluation} (discussing rapid growth of AI coding startups).} Consequently, leading model developers have made a substantial investment at training models to excel at both understanding and writing code—skills which straightforwardly translate to both identifying and exploiting vulnerabilities.\footnote{\textit{See, e.g.}, \citet{cybench}.} Second, cyberattacks occur in a digital environment and do not require access to easily-regulable goods, unlike other forms of AI-enhanced risk: To synthesize a new bioweapon, access to a lab (physically or virtually) is required; to create radiological weapons, one must obtain or manufacture radioactive material. Launching a cyberattack, however, can be done from a laptop. It is therefore easier both to train these capabilities via reinforcement learning and subsequently launch attacks utilizing only access to easily scalable resources (\textit{e.g.} network proxies or virtual machines). Finally, cyberattacks have previously been bottle-necked by human constraints.\footnote{This conclusion follows from the straightforward finding that there are more vulnerabilities available than attackers have the capacity to exploit. \citet{herley2012} examines attacker target selection and concludes that attackers rationally select targets for which a successful exploit is highly likely to translate into economic payoff. \textit{See also} \citet{gray2023}, which examines the internal operations of the Conti ransomware group, including details such as listing open roles with significant requirements for knowledge and training (plausibly suggesting that additional staff would likely translate into expanded offensive operations). Finally, \citet{ablon2017zero} found that vulnerability researchers age out of relevance in approximately three years on average, with the average vulnerability-to-exploit conversion taking approximately three weeks—suggesting a highly skill-intensive process with limited human capital available.} From target identification to the deployment of carefully-crafted social engineering attacks, most aspects of the cyber kill chain previously required involvement of skilled human operators.\footnote{\citet{cyber_kill_chain}.} The ability to substitute AI for human labor threatens to alleviate this bottleneck and produce far more attacks compared to the pre-AI status quo.
\vspace{0.2\baselineskip}

\subsection{The Offense-Defense Balance}
Scholars have discussed the differential impact of increased investment on offense versus defense for many years. The term ``offense-defense balance'' refers to the comparative investment requirements to attain victory, and originates in the strategic and military studies communities.\footnote{\textit{See, e.g.}, \citet{ODbalance} (providing an early formalization of the offense-defense balance).} \citet{garfinkeldafoe} applied the offense-defense balance to cybersecurity and then examined the consequences of scaling investment by both the attacker and defender. They find that when overall investment is low, incremental increases in investment tend to benefit the attacker, while when overall investment is high, incremental increases benefit the defender. This conclusion arises naturally from two characteristics of cybersecurity. First, the attacker can use any vulnerability they uncover that the defender has not identified and fixed, while the defender must work to find all vulnerabilities in their system. Second, for any fixed system, the number of vulnerabilities is finite. The first property means that, at a sufficiently low level of investment, the chance that a given vulnerability identified by the attacker has also been found by the defender is low, while the second property means that a sufficiently high level of investment from the defender can plausibly identify every vulnerability in their system.\footnote{This abstraction, of course, does not fully bear out in reality. For one, not all vulnerabilities are equivalent; some past cyberattacks of note required the use of multiple different exploits. \textit{See} \href{https://www.kaspersky.com/resource-center/definitions/what-is-stuxnet}{Stuxnet}. And, of course, systems are rarely static: New functionality is introduced, vendors roll out new changes to underlying code, and new human employees join the enterprise—each of which can introduce new opportunities that an attacker may take advantage of. Even bugfixes and security refactors can themselves introduce new vulnerabilities.}
\vspace{0.2\baselineskip}

The implications of AI for the cyber offense-defense balance have not gone unnoticed.\footnote{\citet{lohn2022aimakecyberswords}, \citet{lohnimpact}.} Prior works have analyzed how defenders will be able to alter or improve their technical systems with AI—or, conversely, how attackers will be able to utilize AI to identify new forms of vulnerabilities. Andrew \citet{lohnimpact} has provided a thorough examination of how AI's benefits accrue differently to attackers and defenders. He concludes, in part, that AI will likely yield ``a larger number of more complex products to defend'' and potentially accelerate vulnerability discovery, but that it may also assist defenders in designing less vulnerable systems.
\vspace{0.2\baselineskip}

In practice, the relevance of this result is premised on an assumption that companies will \textit{actually} invest in cyberdefense: A trivial conclusion of any offense-defense tradeoff is that some investment from the attacker, with no investment from the defender, will yield an advantage for the attacker. Indeed, cyberdefense is a big business: \$213 billion is projected to be spent in 2025 alone.\footnote{\citet{gartner2025cybersecurity}.} This spending is distributed unevenly, however. For some companies, including many of the largest tech companies in the United States, adherence to cybersecurity best practices is critical to their ongoing business operations. Some of these investments trickle down to end-users of technical services (such as users of Apple's devices or companies which host services on Google's infrastructure), but these end-users also frequently employ enough homegrown software and bespoke infrastructure to require their own investment.
\vspace{0.2\baselineskip}

For many of these leading technical companies, which we will term ``leading-edge organizations,'' this investment is taken as a given. Some of these companies have already begun investing in speculative AI-driven improvements to defensive technology. For one recent example, Google recently used AI to uncover a lurking bug in the SQLite database engine that had gone unnoticed by human reviewers, and subsequently were able to patch that bug before it was released to the world.\footnote{\citet{bigsleep2024naptime}.} More generally, these firms are willing to make expensive investments to reach the frontier of cyberdefense. They often operate \href{https://cloud.google.com/learn/what-is-zero-trust?hl=en}{zero-trust environments}, where even internal technical resources (and personnel) are not assumed to be trustworthy, and some even run \href{https://googleprojectzero.blogspot.com/p/about-project-zero.html}{security research programs} dedicated to identifying and patching zero-day exploits (i.e. \textit{de novo} attacks) in third-party software. The necessity of this investment is not theoretical: Cyberattacks targeting these companies by nation-state actors occur frequently, and compromise means that millions or billions of downstream users are affected.\footnote{\textit{See, e.g.}, \citet{microsoft2024midnight}, \citet{googlechina}, \citet{whatsapp}.}
\vspace{0.2\baselineskip}

For these firms, the precise balance of offensive versus defensive capabilities is highly relevant, as they are willing to make substantial investment in cyberdefense and must assume that their adversaries can likewise spend a substantial amount on offense. Furthermore, they often have the technical expertise to take advantage of technical breakthroughs. It is conceivable that in the near future, teams of automated agents will scour all pieces of software that Google (and similar companies) rely on for vulnerabilities, identifying potential exploits and releasing patches before vulnerable code is ever deployed. For these firms, research identifying the value of cutting-edge techniques like the formal verification of software is highly informative, and can be transformed in short order into internal research efforts.\footnote{\textit{See, e.g.}, \citet{tang2024implications} (discussing the implications of AI for code deobfuscation and zero-day exploit identification), \citet{song2025} (discussing verifiable security), \citet{bradleyrefactor} (proposing automatic rewriting of code from memory-unsafe to memory-safe languages).} Notably, these firms are not infallible and may fall victim to successful attacks periodically,\footnote{Microsoft, for one, was hacked by attackers affiliated with the Chinese government in 2023. The government attributed the success of the attack, in part, to an ``inadequate'' security culture. \textit{See} \citet{exchangereview}.} but they \textit{are} actively engaged in evaluating cyber risk and investing in response to new threat vectors, and moreover have the institutional capacity to improve their defensive posture on a relevant timescale.\footnote{Leading-edge firms still will fail to internalize some societal costs of successful cyberattacks against them,\footnote{\citet{economics}.} they simply have a greater understanding of direct risks and willingness to invest to prevent those harms from materializing.} ``Leading-edge'' refers, therefore, to an organizational willingness to internalize cyber risk, and a high ability to invest in and alter their defensive posture.
\vspace{0.2\baselineskip}

This description characterizes only a small percentage of companies. For most firms, cyberdefense is one priority among many; past industry surveys have shown that less than half of all companies surveyed believe they are ``avoiding major incidents''.\footnote{\citet{cisco2021}. For another example, see World Economic Forum's ``\href{https://www3.weforum.org/docs/WEF_Global_Cybersecurity_Outlook_2024.pdf}{Global Cybersecurity Outlook 2024},'' which in part notes a growing gap between the most- and least-prepared companies: ``The distance between organizations that are cyber resilient enough to thrive and those that are fighting to survive is widening at an alarming rate. As a result, the least capable organizations are perpetually unable to keep up with the curve, falling further behind and threatening the integrity of the entire ecosystem.''} Yet, this discrepancy is rarely captured in the literature on AI's effect on cyber, with some analysis simply noting that defenders may have limited funding or talent,\footnote{\citet{lohnimpact}, 63 (``There is a major shortfall in the number of cyber professionals across all industries. There is also a shortfall in the cyber defense budgets of small organizations, which often include critical infrastructure providers.'').} are impeded by highly bureaucratic processes,\footnote{\textit{See} \citet{hodgson2022managing} for a general examination of organizational responses to substantial cyberattacks. For a particular example, the highly damaging attack on the U.S. government's Office of Personnel Management was the result, in part, of an ``absence of an effective managerial structure'' and ``internal politics and bureaucracy.''} or operate legacy code which is likely to contain vulnerabilities.\footnote{\textit{See, e.g.}, \citet{clarkfamiliarity} noting that legacy code, and code reuse, both correspond to a greater rate of identified vulnerabilities.}
\vspace{0.2\baselineskip}

We argue that these distinctions are more than passing flaws, and instead fundamentally alter the takeaways from the offense-defense balance literature for many companies. Reasoning via offense-defense scaling presumes that defenders internalize risk and can correspondingly invest into improving their defense posture on a relevant timescale.\footnote{Garfinkel and Dafoe wrote the seminal paper on cyber offense-defense scaling, \citet{garfinkeldafoe}. In this paper, they examine outcomes as a function of offensive versus defensive investment, which incorporates the obvious-yet-important result that a defender who does not respond to escalating investment by an attacker will find themselves consistently losing each conflict. Subsequent papers adopt the offense-defense framing and simply compare which benefits an attacker or defender may realize without considering the \textit{propensity} of those firms to invest in obtaining those benefits.} We argue that a large share of companies do not meet these criteria and consequently are exposed to risk that is not effectively characterized by the offense-defense balance. We term these companies ``trailing-edge organizations.'' These firms invest only minimally in cyber in the status quo and are likely to be unprepared for two interlocking changes brought about by widespread use of AI. First, AI will drive a significant decrease in the marginal cost of launching a cyberattack. Second, attackers will be able to utilize AI to identify more vulnerabilities more quickly, in part because of technical processes and systems that these organizations rely on.

\vspace{0.2\baselineskip}

In Section 2, we begin by exploring the status quo, and conclude that cyberattacks are limited by the economics of human labor and exploit development, not a dearth of vulnerable targets. Next, in Section 3, we discuss the near-term implications of advances in AI. In particular, we argue that advances will yield both a substantial reduction in the marginal cost of launching a cyberattack (Section 3.1), and novel technical methods for developing exploits (Section 3.2). These shifts will subject trailing-edge organizations to a surge in attacks, yielding both economic consequences for the firm and major negative externalities for consumers and societies. Finally, in Section 4, we examine potential counterarguments including benefits to defenders, before providing a small set of recommendations for both trailing-edge organizations (to begin improving their defensive posture) and governments (to encourage defensive modernization) by firms.

\section{The Status Quo}
If leading-edge organizations are characterized by well-staffed cybersecurity teams that implement (or, in some cases, define) best practices for cyberdefense, trailing-edge organizations are characterized by the opposite. These firms often do not assign sufficient funding or staffing to comply with existing best practices, let alone new developments that are the result of improved AI systems. Many trailing-edge organizations are also subject to operational, technical, and procedural debt that is both costly to address and does not likewise straightforwardly benefit from advancements in AI.
\vspace{0.2\baselineskip}

For these companies, cybersecurity is not a top corporate priority by default. Industry surveys and other research suggest a large cohort of companies regularly fail to prioritize deploying security patches,\footnote{\citet{clones}.} do not deploy standard technologies like intrusion detection software, and fail to automate key processes, guaranteeing that humans must remain in the loop for time-sensitive operations.\footnote{\citet{cisco2021}.} For example, the UK government found that only 32\% of UK businesses had a policy to apply security updates within two weeks of release, only 40\% had adopted two-factor authentication, and only 19\% provided any cybersecurity training to staff within the last year.\footnote{\citet{dsit2025cyber}.} Instead of investing proactively, these companies generally tend to dramatically increase their spending following a successful cyberattack against them.\footnote{\citet{ibm2025cost}.} A plausible explanation of this phenomenon is that companies invest based on prior estimation of expected damages of a cyberattack. From these statistics, then, it appears that their leadership simply does not believe that they are likely to be attacked, or alternatively believes that even if they are attacked, resultant damages will be low.\footnote{These assumptions are likely not correct, even today. A single data breach can have ``devastating'' impacts on a company's business, \citep{huang2023devastating}, and substantial additional costs to consumers or society. The company is rarely liable for these additional costs (e.g. leaked personal email addresses or credit card numbers), however, so companies may rationally choose not to internalize those risks, \citep{economics}.}
\vspace{0.2\baselineskip}

This is not, on the whole, an incorrect assumption for the world today. Indeed, attackers are likely insufficiently economically motivated to launch many potential attacks. Literature has considered the incentives of attackers through an economic lens for decades.\footnote{\citet{investmentStrategies}.} A simplified understanding would suggest that when the marginal dollar invested into launching a cyberattack yields less than a dollar of expected return, the attack is not launched. This model is alluringly simple—but to fully capture an attacker's cost-benefit analysis, it must incorporate several additional aspects of how cyberattacks occur in practice. First, in the world prior to AI, launching an attack requires the involvement of a human at many key steps. Second, the total amount of human capital available for launching attacks is limited. And third, the real-world threat of criminal sanctions likely deters attacks beyond what a purely economic model would predict.
\vspace{0.2\baselineskip}

In the status quo, humans are required for many stages of an attack. The cyber kill chain refers to the sequence of steps an attacker takes in identifying a target, crafting an exploit, and launching the attack itself.\footnote{\citet{cyber_kill_chain}.} These steps involve many stages which require human judgment and ingenuity. From identifying the correct target to lateral motion within a heterogeneous technical environment, few steps could be taken in an automated fashion if their success was to be guaranteed—at least, prior to the advent of advanced AI systems.\footnote{An obvious exception to this is self-propagating exploits, or worms, which rely on automatically spreading to other infrastructure that has the same vulnerability, such as the \href{https://www.fbi.gov/news/stories/morris-worm-30-years-since-first-major-attack-on-internet-110218}{Morris Worm}. See \citet{cybench} for an argument that AI can assist at multiple stages of the kill chain.}
\vspace{0.2\baselineskip}

If human bandwidth is the limiting factor, then the key reagent is how many trained attackers there are. There are many—but nowhere near as many as there are vulnerable companies. In part, this is due to the rise of lucrative software jobs, which reward many of the same skills as cyberoffense. If a job at a tech company guarantees you a comfortable lifestyle, then many rational actors will choose to abandon their unlawful activities. This is not universal: There are some markets where there is not a substantial domestic tech industry, making cybercrime a more attractive option for individuals with the requisite skills. But again, given the substantial scope of vulnerable companies suggested by the evidence presented above, it is highly likely that the number of attacks launched was directly limited by the number of humans with the appropriate training.
\vspace{0.2\baselineskip}

Then, of course, there is the effect of the nation-state itself. A cyberattack is rarely a victimless crime, and law enforcement has repeatedly tracked down and arrested individuals responsible for particularly damaging attacks.\footnote{For a recent, high-profile example, see the Department of Justice's recent apprehension of an alleged Chinese cybercriminal while he was traveling in Italy. \citet{hafnium}.} Here, three forces combine to limit the scope of attacks that are launched. First, the intelligence community has been able to identify the perpetrators of many cyberattacks with a high degree of confidence due in large part to human signals present in those attacks (\textit{e.g.} patterns of exploit design, reuse of malware, or locations of where the attack was launched from.)\footnote{For an example of the steps of this analysis, see \citetalias{criminal}.} Second, international cooperation via extradition treaties, plus the long arm of the United States Justice Department, means that being overseas from your target is not a categorical bar to facing criminal consequences. And finally, nations often lean on domestic cyberattackers to ensure that their attacks do not create negative geopolitical blowback\footnote{\textit{See, e.g.}, \citet{darkside} (detailing dissatisfaction from the Russian government with a domestic hacking group allegedly responsible for the Colonial pipeline attack).}—as is the case for China and Russia, whose large-scale attacks are thought to have the blessing of the central government.\footnote{\textit{See, e.g.}, \citet{benner2020equifax}.}
\vspace{0.2\baselineskip}

These factors combine to produce a world, prior to the widespread use of AI, that is remarkably \textit{lacking} in damaging cyberattacks. Most organizations have been able to escape from the effects of highly compromising attacks like WannaCry. This is, in a sense, a form of ``security by obscurity,'' insofar as organizations rely on being a low-profile target without obvious and sufficient economic incentives for attackers to try their hand.\footnote{For the counterargument that attackers instead forbear from attacking most targets to avoid attracting nation-state attention, \textit{see infra} note 57 and accompanying text.}
\vspace{0.2\baselineskip}

\subsection{Human-Scale Failures}
When attacks do occur against these companies, they frequently succeed not because of a brilliant technical breakthrough (that is, the discovery of a novel technical vulnerability in a piece of field-hardened code), but instead because of failures of human systems.\footnote{\textit{See, e.g.}, \citet{song2025} (slides 77-79), discussing how many attacks begin with human failures (focusing in particular on social engineering attacks). See also \citet{huang2025beyond}, discussing human factors of cyber risk.} We highlight two categories of failures that are common, and have, in the past, enabled highly-damaging attacks. First, substantial and regular delays in deployment of security patches. Second, technical systems which place high trust in and rely on individual human operators—whether for the purposes of institutional knowledge, human sign-off, or merely due to granting them broad technical permissions which are not required for job responsibilities.
\vspace{0.2\baselineskip}

The 2017 WannaCry attack affected hundreds of thousands of computers worldwide and caused more than \$4 billion in economic losses. The exploit would infect a single computer, encrypt the files on the hard drive, demand a ransom, and then spread over the network to other machines. It was \href{https://www.npr.org/sections/thetwo-way/2017/12/19/571854614/u-s-says-north-korea-directly-responsible-for-wannacry-ransomware-attack}{attributed to North Korea}, though the rationale for the attack is still unclear: Despite affecting an enormous number of machines, only approximately \$250,000 was sent to the Bitcoin address in question.\footnote{The attack's impact was dramatically limited due to the actions of an unaffiliated cybersecurity expert and once-blackhat hacker, Marcus Hutchins, whose \href{https://www.wired.com/story/confessions-marcus-hutchins-hacker-who-saved-the-internet/}{story} is worth reading in its entirety.} Yet, the attack itself provides a template for how delays in patching vulnerabilities can lead to widespread economic costs.
\vspace{0.2\baselineskip}

WannaCry relied on a zero-day exploit, known as EternalBlue, for Microsoft Windows. The exploit was known to (and possibly used by) the National Security Agency (NSA) for years in advance of the attack. It was later \href{https://tribune.com.pk/story/1423609/shadow-brokers-threaten-release-windows-10-hacking-tools/}{leaked}, however, by a group of third-party hackers, and the NSA was then forced to warn Microsoft about EternalBlue's existence.\footnote{A reasonable conclusion from this story is that government actors should not harbor exploits in software made or relied upon by American companies. However, our argument does not rely on this point, and it will not be discussed further.} Microsoft released a \href{https://learn.microsoft.com/en-us/security-updates/SecurityBulletins/2017/ms17-010}{patch} approximately a month before a proof-of-concept of the exploit became publicly available, and two months before the WannaCry attack began.\footnote{\citetalias{criminal}.} A critical period therefore existed between the release of the patch and the first public demonstration of the exploit by a security researcher: If a system was patched in this window, then it was immune to any resultant malware; if it was not, then malware relying on the vulnerability—like WannaCry—could spread. Thus, classifying WannaCry as reliant upon a zero-day exploit elides the true reason for the vulnerability: substantial delays in patch deployment among downstream users. This type of failure affects primarily trailing-edge companies—and the risk of attack increases with each day a patch is delayed, as AI enables faster patch development and scales target identification and reconnaissance.\footnote{\textit{See infra}, section 3.2.}
\vspace{0.2\baselineskip}

A two-month delay between patch release and deployment is, unfortunately, not the low-water mark for cyberdefense among critical enterprises. In 2017, the consumer credit reporting agency Equifax failed to apply a critical security patch for five months, resulting in a vulnerability that Chinese state actors exploited to exfiltrate records relating to approximately 148 million Americans.\footnote{\citetalias{houseequifax}, 2–4.} The vulnerability lay in a piece of open-source software, Apache Struts, used by Equifax to build web applications in Java. The vulnerability, \href{https://nvd.nist.gov/vuln/detail/cve-2017-5638}{CVE-2017-5638}, was announced by Apache on March 6, 2017. The Department of Homeland Security, recognizing the severity of the vulnerability, alerted Equifax to the advisory, and notice was then widely broadcast within the company. The patching process began immediately and concluded a few days later.\footnote{\citetalias{houseequifax}, 2. A primary cause of these failures is perhaps predictable: Human systems, human squabbling, and human scarcity. Equifax's operational and security responsibilities within the IT department were split between two separate executives (the CSO and CIO), those two groups of IT professionals communicated in a highly inefficient manner, and company leadership thought of cybersecurity as one business objective among many.} Their work was incomplete, though: A web application continued running the older version of Struts, nestled in a legacy portion of Equifax's technical infrastructure. Two months later, on May 13, the cyberattack began. It took Equifax an additional 76 days—until July 30—to identify the unusual traffic and shut down the vulnerable application.
\vspace{0.2\baselineskip}

The oversight of the single server could be attributed to a forgivable human failure—it was, after all, just one system. If the attackers had not been sufficiently motivated, it is possible it could have gone unnoticed and eventually been replaced or upgraded as part of routine technical maintenance. Yet, this failure is representative of broader issues within Equifax, as evidenced by congressional testimony following the hack. The final report noted, ``[Equifax]'s lack of knowledge about the software used within its legacy IT environment was a key factor leading to [the hack]. Equifax's Patch Management Policy relied on its employees to know the source and version of all software running on a certain application in order to manually initiate the patching process.''\footnote{\citetalias{houseequifax}, 74.} This, needless to say, does not scale. An employee can forget about a server that they have not worked on in a while, while an automated registry (like any major orchestration solution would automatically provide\footnote{\textit{E.g.} \href{https://kubernetes.io/}{Kubernetes}.}) does not lose track of machines. But Equifax is not some no-name small business; it had an executive responsible for cybersecurity, a staffed IT department, regular audits, and a separate security engineer role—so why did they not recognize these flaws?
\vspace{0.2\baselineskip}

As the congressional testimony would later reveal, they did. Years prior to the attack, they conducted an audit of their patch management processes, which turned up eight key deficiencies, along with recommendations to address them.\footnote{\citetalias{houseequifax}, 69.} Those recommendations included, ``implement[ing] automated patching tools,'' ``improv[ing] IT asset management controls to ensure a[n] . . . accurate inventory of all IT assets is available,'' and ``creat[ing] a centralized patch and exception process.''\footnote{\citetalias{houseequifax}, 69.} Fatefully, these recommendations went unheeded, despite theoretical due dates for each that would have likely resulted in them being mostly or entirely implemented prior to the Chinese intrusion.\footnote{\citetalias{houseequifax}, 69–70.}
\vspace{0.2\baselineskip}

Prior to the attack, it may have been difficult for an outside observer to guess that Equifax's security practices were so lacking, which is a testament to how common these flawed practices are among even well-known businesses. It is clear that Equifax should have been on notice of the risk of major cyberattacks, given that their direct competitor Experian had been subject to a similarly-sized attack only a few years prior. And Equifax had identified flaws in their security posture years prior, as noted above, but failed to remediate those flaws in a timely manner. This, we think, captures the struggle of the typical trailing-edge organization: In some cases, they may not believe a cyberattack is likely or even possible; but if they do, organizational gridlock and lack of human capacity severely hamper their ability to quickly implement changes.\footnote{In Equifax's case, the answer was clearly both. Operational and security responsibilities within the IT department were split between two separate executives (the CSO and CIO), those two groups of IT professionals communicated in a highly inefficient manner, and company leadership thought of cybersecurity as one business objective among many. Furthermore, periodic audits of Equifax's data security practices revealed shocking levels of incompetence, with one audit rating the company's efforts a ``zero out of ten.''} Critically, this inertia will be differentially riskier in the near future, as attackers take advantage of AI's improved capabilities to develop attacks more quickly and against more targets.\footnote{\textit{See infra}, section 3.2.}
\vspace{0.2\baselineskip}

In the status quo, many organizations will be able to scrape by with this defensive posture. A few will suffer enormously damaging attacks, but they will recover, attempt to implement sweeping reforms to address their deficiencies, and carry on.\footnote{In Equifax's case, that reform involved the departure of their CEO, CSO, and CIO; the payout of at least \$380 million to compensate users; and the wholesale reinvention of their security division. \textit{See} \href{https://www.equifaxbreachsettlement.com/faq}{Equifax Data Breach Settlement}, \citetalias{houseequifax}, 48–49.} Most, however, will simply escape notice—and the lack of a damaging attack in one year justifies no increase to the organization's cybersecurity budget in the next. Yet, as we argue next, many basic assumptions about the status quo are likely to change in the coming years due to the adoption of widespread AI. Therefore, for trailing-edge organizations, the question is not whether AI will benefit offense or defense more, because many of these organizations already operate with wildly deficient cybersecurity departments. The question is instead how much \textit{additional} risk these firms face due to their deficient security practices—and how they can begin to close the gap with their leading-edge peers.

\section{The Shifting Threat Landscape}
We argue that two factors combine to make trailing-edge organizations markedly more likely to suffer a damaging cyberattack in a near-future world with advanced AI systems.\footnote{There are several preconditions to this world arriving: Continued scaling of AI capabilities, for one, and the ability for malicious actors to gain access to misaligned AIs which will happily produce code for an exploit (or, alternatively, sufficiently effective prompt hacking techniques to make a safe model usable for illicit purposes.) Though these assumptions may be contested, they are the basis for many predictions within the AI safety community, and we do not examine them further here. \textit{See, e.g.}, \citetalias{ai2027}  (predicting widespread autonomous attacks by February 2027) and \citet{ncsc} (predicting increased cyber threats from both state and non-state actors due to AI).} First, the previously-limiting factors on the number of attacks launched will recede: More actors will be able to launch more attacks for less money. Second, attacks will be able to make use of vulnerabilities more quickly: Existing practices for responsible disclosure, reliance on third-party software, and open-source code all present opportunities for motivated attackers to launch attacks before trailing-edge organizations are able to mount an effective defense.

\subsection{The Decreased Cost of the Marginal Attack}
Three forces combine to change the economics of launching a marginal cyberattack. The ability to delegate operational control to an AI system while the attack is ongoing will substantially reduce the requirement of human supervision and control. The ability to launch attacks without sophisticated technical knowledge will expand the pool of potential attackers. Last, the ability to disguise the source of an attack will remove many institutional and geopolitical checks on attackers that would have stopped them previously. In combination, these promise to make many organizations viable targets when previously it was simply not worth an attacker's time to pursue them.
\vspace{0.2\baselineskip}

Widespread use of advanced AI removes the human bottleneck on launching attacks by allowing attackers to delegate control of the attack to an AI system.\footnote{\textit{See, e.g.}, \citet{rubin2025agentic} and \citet{cvebench}.} This is of particular concern when large numbers of targets are made available at once—for example, when large batches of stolen user credentials are posted to the dark web, or a vulnerability is found in a widely-used product. \href{https://arxiv.org/abs/2503.17332}{Recent benchmarks} have shown that frontier LLMs can already, in a non-negligible proportion of cases, completely autonomously compromise such systems without any further human input once the attacker has identified the vulnerable endpoint and given a high-level CVE description. Of course, merely gaining access to a system is not the damaging part of an attack, but the ``dwell time'' needed for an attacker to cause serious damage is substantially reduced.\footnote{Defined as the duration a malicious actor remains undetected within a system after successfully breaching it.} A cybersecurity company recently demonstrated data exfiltration using an agentic AI framework taking orders of magnitude less time than similar pre-AI attacks.\footnote{\citet{girt}; \textit{see also} their more general assessment and threat modeling of agentic AI attack risks \citep{rubin2025agentic}.} Once sensitive information is obtained, AI allows attackers to weaponize it faster and more effectively, identifying the key parts needed for further compromise of the organization or the most valuable assets to steal or destroy.
\vspace{0.2\baselineskip}

Second, AI threatens to dramatically expand the pool of individuals that are capable of launching attacks. Before, a foundational knowledge of computer science was a bare minimum, and obtaining that knowledge required a substantial time investment, whether via studying textbooks or obtaining a degree. Now, an AI can coach a user through the steps required to set up cloud infrastructure, conduct reconnaissance, or begin using \href{https://www.metasploit.com/}{Metasploit}; it may soon, as noted above, be able to fully offload \textit{all} human work, relegating the attacker to the managerial role of simply describing the AI's targets and objectives.
\vspace{0.2\baselineskip}

Third, near-future AI systems threaten to obfuscate many of the markers traditionally used by law enforcement for identifying the source of an attack, and hence, a greater chance that attackers will escape consequences. \citet{attribution} describe cyber attribution as an art which relies on factors as varied as digital forensics, strategic assessment, behavioral markers, and intuition. In their analysis, attribution is not a binary yes-or-no, but instead a spectrum from high- to low-confidence. Experienced cyberanalysts look for malware reuse and shared tools—markers which can be reliable indicators of attacker identity because many perpetrators are repeat players (\textit{e.g.} criminal groups or state actors).\footnote{\citet{complexity} and \citet{evasive}.} Furthermore, metadata like attack timing or file naming can give a clue to the location of attackers. Finally, a geopolitical lens can suggest which nations stand to benefit the most from an attack on a particular target.
\vspace{0.2\baselineskip}

AI-driven attacks threaten to disrupt these analytical pillars: AI can innovate new exploits, minimizing the information gleaned from digital forensics; it can operate around the clock and without leaking any behavioral insights about its operators (in part because it will not know who is operating it); and because non-state-actors will newly have the ability to launch attacks, geopolitical analysis will be less revealing. Defenders will also benefit from improved telemetry, and in time, new analytical methods will undoubtedly be developed, but in the short term, attackers are likely to benefit from increased obscurity. In this future, we believe that it is highly unlikely that attackers will cooperatively restrain from attacking firms to avoid drawing nation-state attention due to the difficulty of both estimating damages and coordinating without centralization. Consequently, a lessened ability to attribute attacks will likely translate directly into more attackers being willing to take their chances.
\vspace{0.2\baselineskip}

In the past, the limited number of capable attackers and the potential real-world consequences of attacks meant that many trailing-edge companies were able to escape notice even with sub-par defensive practices—though when ``security by obscurity'' failed, investment in better cybersecurity practices followed immediately after.\footnote{This investment is not only too late to stop the first attack, but may be less effective on the whole: \citet{proactive} examine cost-effectiveness of proactive versus reactive investment in the healthcare sector, and conclude that voluntary, proactive investment is the strongest indicator of positive security outcomes.} In the near future, we predict that advances in AI will mean many fewer firms will escape notice by default. Improvement and diffusion of AI mean that attackers will no longer need to be in the loop for many decisions, lower-skill attackers will benefit from skill uplift, and nation-states will struggle to identify perpetrators of attacks without reliable attack metadata.\footnote{A straightforward counterargument attributes the lack of attacks to the forbearance of attackers, and in particular, a desire to avoid causing so much damage that nation-states have no choice but to respond. We believe two key factors weigh against this understanding. First, with lower marginal costs for attack, more individuals will attempt attacks, making strategic non-attack a substantially more difficult coordination problem. Second, attackers cannot perfectly estimate damages \textit{a priori}. For example, in the 2021 Colonial Pipeline ransomware attack, the alleged attackers ended up disclaiming an intention to cause widespread fuel shortages and damages—but were unable to predict those consequences ahead of time \citep{darkside}.} This, however, is not the end of the trouble for trailing-edge corporations. The same increases in AI capabilities will enable attackers to craft attacks more quickly against more targets, meaning that the goal is shifting for these companies: Not only must they harden their systems to the attacks of the past, but they must be ready for a new wave of greater-threat attacks.

\subsection{Technical Threats from Common Practices}
\vspace{-0.15\baselineskip}
Thus far, our discussion has centered on how advances in AI affect the economic dimensions of cyberattacks and cyberdefense. The implications for technical aspects of attacks, however, are also striking. We argue that many of the long-established practices of the software industry, from responsible disclosure to a reliance on open-source software, pose heightened risks with AI-enabled cyberattacks. We highlight three causes: First, AI will be able to more effectively translate vulnerability disclosures into working exploits, dramatically shortening the time between disclosure and attacks on unpatched systems. Second, given that the existence of one vulnerability in a system often implies other vulnerabilities of similar types hidden elsewhere,\footnote{For example, the discovery of a vulnerability caused by a lack of memory-safety suggests that similar vulnerabilities may exist in the same or neighboring code; an attack based on failures in parsing untrusted attachments may indicate other vulnerabilities of the same form.} AI will likely enable attackers to rapidly develop multiple, redundant exploits of the same target—making attacks resilient against single patches, especially if the code of the fix is publicly accessible. Third, AI will dramatically accelerate the process of target identification by analyzing data like port scans in an automated fashion and correlating it with other public records, shortening the time vulnerable systems can remain undiscovered.
\vspace{0.2\baselineskip}

Each of these attributes means that the threat landscape will not simply be the attacks of yesterday launched by new attackers; the threats themselves will arrive more quickly, be more resilient to patching, and take aim at a wider array of targets. The consequences for those trailing-edge organizations that fail to adhere to cyber best practices today are substantial—not only must they modernize their practices and overcome organizational inertia, they must surpass the previous high-water mark if they are to effectively defend against novel attacks.
\vspace{0.2\baselineskip}

Current AI capabilities already provide significant uplift to attackers at multiple stages of the kill chain and we expect this gap to widen in the near-term future.\footnote{\citet{girt}; \textit{see also} \href{https://x.com/dawnsongtweets/status/1935436249923469594}{this thread on X} by Dawn Song.} In particular, attackers will benefit from defenders' slow and bureaucratic patch deployment processes; use of legacy, custom-built code; and high reliance on human actors. In this section, we identify how recent AI improvements change the arithmetic for patch deployment timelines, threaten to uncover additional similar exploits in the same code, and scale up target identification.
\vspace{0.2\baselineskip}

First, it becomes easier for less sophisticated actors to convert theoretical knowledge of a vulnerability’s existence into code which allows them to attack affected systems. When security issues are discovered, a process known as responsible disclosure typically occurs, yielding an entry in the Common Vulnerabilities and Exposures (CVE) system.\footnote{The responsible disclosure process takes place in a number of key stages, which we summarize here:
\begin{itemize}[leftmargin=3em]
\item Discovery and notification: a potential security issue with a particular piece of software or hardware is identified by a third party, typically a security researcher. They make contact with the vendor of the affected systems and provide technical details showing that the vulnerability exists.
\item Verification and fix: the vendor confirms the vulnerability exists, assesses its severity and impact, and determines priority for fixing it. If the vendor decides that the bug is not a security issue, the details are typically published immediately and publicly. The vendor identifies the affected versions, creates a fix, tests it thoroughly to ensure it doesn't break existing functionality, and prepares adistribution mechanisms.
\item Announcement: The vendor releases the patch publicly and publishes a security advisory, typically in the form of an entry in the Common Vulnerabilities and Exposures (CVE) system. This advisory will contain a unique ID, a list of the affected versions, a brief technical description of the issue (sufficient for customers to determine whether it affects their use-case) and an assessment of severity. It does not typically contain a detailed technical explanation of the issue or proof-of-concept code (in part to raise the bar for attackers to exploit it) although the researcher or others may also simultaneously publish technical details about the vulnerability.
\item Organizational remediation: potentially affected users of the relevant products review their deployments to determine if they are using the vulnerable versions, and if so whether they actually use the feature in question, whether specifics of their use-case stop it being exploited, and how urgent it is to implement a fix vs. workaround. If it is deemed necessary to take the fix, they will then need to create and implement a plan to update their vulnerable component, test the integration of the fix, and then roll out the new version without causing excessive business impact.
\end{itemize}
} CVE descriptions typically include a technical description of the security issue and a list of affected versions. For organizations which deploy the affected software, this information is typically sufficient to identify whether or not they must apply a patch. Historically, this disclosure process did not provide an attacker with enough detail for them to reproduce the patch; this delay is why many cyberattacks begin only after researchers release a working proof-of-concept following the patch process.\footnote{Google Project Zero \href{https://googleprojectzero.blogspot.com/p/vulnerability-disclosure-faq.html}{explicitly states} this assumption when explaining why they have a 90 day deadline for disclosure of vulnerabilities they identify.} However, this claim may no longer be true: A security researcher recently described how they were able to exploit a recently-found security hole in Erlang’s SSH library by asking GPT-4 to analyze the relevant patches and create a proof-of-concept to attack vulnerable systems.\footnote{\citet{claburn2025ai}. \textit{See also} \citet{automatic}.} One of the authors of this piece had a similar experience while creating CVE-Bench:\footnote{\citet{cvebench}.} He was able to successfully exploit several critical- or high-severity CVEs within a matter of hours, relying only on Claude Sonnet 3.5, the published CVE description, and access to the commit history of the repository.
\vspace{0.2\baselineskip}

A patch for a specific issue may also signal to attackers that the targeted software is vulnerable to a general class of attacks, which may then be identified and exploited before the vendor knows about them all. In recent weeks, Microsoft announced a patch for a vulnerability in SharePoint, a tool for organizational file-sharing, that was not believed to be exploited in the wild.\footnote{\citet{hackersharepoint}.} Days later, however, Chinese nation-state attackers began launching clusters of attacks premised on a similar, unpatched vulnerability in the same software.\footnote{\textit{See} \citet{hackersharepoint}, with associated CVEs \href{https://www.cve.org/CVERecord?id=CVE-2025-53770}{CVE-2025-53770} and \href{https://www.cve.org/CVERecord?id=CVE-2025-53771}{CVE-2025-53771}.} It is likely this pattern will repeat itself, especially for large, legacy applications that are only infrequently updated. A particular choice of language or framework, or repeated software design patterns, can mean that there are clusters of adjacent vulnerabilities scattered across the same codebase.\footnote{\citet{hernan2023azure}.}
\vspace{0.2\baselineskip}

Relatedly, it becomes easier to scale analysis of code added to open-source repositories for new security issues, and extrapolate from in-progress fixes to identify existing unmitigated vulnerabilities.\footnote{Closed-source vendors are not immune to this, though it may form a defense; APTs such as Midnight Blizzard have gained access to commercially sensitive source-code repositories \citep{msrc2024midnight}.} Projects which are used in millions of systems worldwide are often maintained by a very small number of volunteers or underfunded developers with a minimal budget for security review.\footnote{\textit{See} \citet{cisa2024}.} Since attacks on complex systems often only require a small number of failures in defense, adversaries with a large inference budget can convert this into a more detailed understanding of many small pieces than the individual maintainers themselves might have. Furthermore, each of these libraries and components is generally designed and tested in isolation, making their own assumptions about data validation and formatting, memory management, or other shared resources.\footnote{For example \href{https://nvd.nist.gov/vuln/detail/cve-2015-0336}{CVE-2015-0336}, caused by a confusion between the “type” of certain objects, and \href{https://nvd.nist.gov/vuln/detail/CVE-2000-0884}{CVE-2000-0884}, where a server made incorrect assumptions about the characters permitted in a URL.} A violation of thread-safety might only become exploitable when combined with a specific mismatch in error handling of a different library and the particular resource management pattern of a third. AI-powered analysis could potentially trace these complicated assumptions across the entire software stack, identifying exploitable combinations that no human reviewer could feasibly discover.
\vspace{0.2\baselineskip}

Third, AI makes it easier to identify vulnerable systems. Although using automated cyber threat intelligence collection to find potential targets is not a new development, AI substantially improves its capabilities across several dimensions. \citet{google2025} highlights the use of their Gemini AI offering by APTs\footnote{APT stands for an advanced persistent threat, meaning a sophisticated and long-standing group of attackers, typically sponsored by a state or experienced criminal organization.} from China, Iran, and North Korea for a variety of target reconnaissance activities, including gathering details about the attack surfaces, network configurations and system architecture of specific targets. Undirected attacks are also possible: Applications typically use a consistent port when running on a particular server, and \href{https://github.com/robertdavidgraham/masscan}{modern network tooling} can scan the entire IPv4 space to find these systems in just a few minutes. Once potential targets have been found, AI systems could probe them to detect differences in behavior\footnote{These could be publicly known—either because the feature set is different or the responses explicitly allow version identification—or more subtle, such as gleaning information from error handling or attempting to hit known bugs.} and determine whether the systems are using exploitable versions of their software.\footnote{\citet{electronics}.}
\vspace{0.2\baselineskip}

Even when a company makes a substantial investment in security and develops a well-oiled remediation process, their systems may still be vulnerable if they have become dependent on a less capable third-party software provider. The \href{https://www.heartbleed.com/}{Heartbleed} vulnerability (\href{https://www.cisa.gov/news-events/alerts/2014/04/08/openssl-heartbleed-vulnerability-cve-2014-0160}{CVE-2014-0160}) was a 2014 OpenSSL bug which allowed attackers to easily steal extremely sensitive information from servers.\footnote{OpenSSL is a widely-used, open-source cryptographic library that provides the fundamental building blocks for secure communications on the internet.} It had a simple fix, meaning that users of affected web servers, browsers, and VPNs were able to quickly upgrade and secure their systems. However, those products were themselves used by the vendors of a wide variety of products with a web interface. Users of those products had to wait for the maker of their industrial control system, network firewall manager, etc., to take the new version of the web server, rebuild their firmware, perform extensive testing, and roll out the final update.\footnote{\citet{embedded}.} This process took weeks, or, in cases where the initial vendor had gone out of business, never happened at all. With that said, if the alternative to relying on third-parties with suboptimal security practices is using internally-created and maintained services, this may still be a better option if the company is not willing or able to make sufficient investment in keeping those internal services up-to-date and secure.\vspace{0.2\baselineskip}

The technical consequences of AI's increased capabilities, therefore, are substantial, but many observers have noted that advancements in AI capabilities will also provide substantial uplift in \textit{defensive} capabilities. In the next section, we briefly summarize these defensive benefits, examine why trailing-edge organizations will still be subject to heightened risk, and then present a prediction of what the near-future looks like for trailing-edge organizations absent rapid action. We conclude with a set of recommendations for trailing-edge organizations and governments to consider as part of a holistic response.

\section{Consequences and Recommendations}
In this section, we examine a number of counterarguments to the grim picture we have painted for trailing-edge organizations. We begin by examining the consequences for defensive capabilities due to improved AI capabilities, and whether those capabilities are likely to be realized by trailing-edge organizations. We then examine likely outcomes in a world without any substantial interventions—in essence, examining what \textit{actually} happens if many organizations are exposed to dramatically heightened risks of cyberattacks, and what response will play out at a societal level.

\subsection{Defensive Consequences}
Some observers have predicted that substantial benefits will accrue to defenders as AI's progress continues, potentially even outpacing corresponding improvements to offensive capabilities.\footnote{\citet{lohnimpact}.} Yet, as we have argued, likening every company with a security team to Google elides key distinctions in how willing companies are to spend on cyberdefense, and how many human-scale vulnerabilities are peppered throughout their organization and technical systems. We now consider how these defensive opportunities translate from leading-edge to trailing-edge companies.
\vspace{0.2\baselineskip}

Current and near-future AI capabilities will provide assistance to defenders, and we see this being consequential in several areas. First, AI analysis and code generation may make it easier for vendors to build and test fixes for security issues.\footnote{\citet{huang2025aimodelshelpproduce}. \textit{See also} DARPA's recent ``\href{https://aicyberchallenge.com/}{AI Cyber Challenge},'' where teams competed to use AI to identify and patch synthetic vulnerabilities in a range of open-source projects.} Both Google and Meta have demonstrated success in using LLMs to generate fixes for vulnerabilities found during runtime fuzzing.\footnote{\citet{nowakowski2024ai} and \citet{byun2025autopatchbench}.} This class of bugs—particularly in memory-unsafe languages such as C and C++—frequently results in severe security vulnerabilities such as remote code execution or memory disclosure.\footnote{\citet{ifp}.} However, the speed of producing fixes does not currently act as a bottleneck to securing most systems from vulnerabilities,\footnote{For example, Google's Project Zero \href{https://googleprojectzero.blogspot.com/p/vulnerability-disclosure-faq.html}{claims} that 96.9\% of zero-day issues they identified were fixed within the 90-day window prior to public disclosure (including a potential one-time extension for high severity issues).} and while it may allow the shortening of disclosure time for vendors (which shrinks the window for attackers to use zero-days against leading-edge targets), many attacks succeed long after a fix has been prepared due to companies delaying the application of security patches, as we have argued above.
\vspace{0.2\baselineskip}

Second, AI systems provide an additional, cheap layer of pre-deployment oversight. A crucial defensive asymmetry in cyber is the defender’s control over the landscape itself in their choice of systems to create and code to deploy.\footnote{\citet{lohnimpact}.} In the long run, we expect a significant shift towards defense here because AI-powered code- and architecture-review will reduce the number of bugs per line of code deployed to production.\footnote{Although this does not necessarily imply there will be less total bugs; we do not take a position on whether the increased rate of generating code would outweigh each individual part being more secure.} For the time being, however, most of the code in production has already been written and frontier AI systems struggle substantially more with refactoring old code than writing new code.\footnote{\citet{llmassist}.} The defensive advantage here is therefore small but will increase as AI systems are involved in the design, development, and deployment of features from the beginning.
\vspace{0.2\baselineskip}

Third, AI may make it easier to maintain an accurate inventory of digital assets and understand when systems are affected by particular CVEs.\footnote{\citet{electronics2}.} By analyzing system logs and network traffic, AI can automatically discover and catalogue devices, software components, and dependencies. This, conceivably, would have identified Equifax's vulnerable legacy servers without relying on human recollection. This discovery process can operate in real-time, with AI detecting configuration changes and software updates and performing analysis without manual intervention. When new CVEs are published, this real-time asset inventory makes it much easier to identify affected systems and crucially, it reduces the risk that old, unpatched, and forgotten systems remain as entry points for attackers—attributes which likely would have revealed the unpatched server to Equifax's security teams. Conversely, AI can also help rule out CVEs that don't actually affect particular deployments, freeing the security team to focus on other pressing issues. For example, if a system uses a vulnerable version of software but doesn't enable the specific \textit{feature} that contains the security vulnerability, AI systems can indicate that no action is required.
\vspace{0.2\baselineskip}

Finally, modern AI and ML methods offer a rich and powerful toolkit for identifying and understanding suspicious activities within corporate networks.\footnote{\citet{intrusion}.} Critical parts of the attack chain—including the initial authentication and connection to the network, search and analysis of company assets, deletion or encryption of files, and exfiltration of data outside of the network—may superficially resemble legitimate or typical traffic for the organization, but sufficiently intelligent analysis would reveal its malicious nature. At the moment, the vast scale and complexity of network traffic analysis hinders the reliability of current solutions in this space, but \citet{saleh} have demonstrated that models have the potential to identify cyberattacks with a high degree of precision and accuracy. Attackers will undoubtedly develop new techniques to evade notice, but we expect defenders to eventually achieve a decisive advantage. Defenders have intelligent control over each part of their network infrastructure\footnote{\textit{I.e.}, they can in theory analyze network traffic and update filtering and blacklists in real-time based on their assessment of whether traffic is legitimate.} and can achieve comprehensive observability with adequate investment, while attackers can only evade notice for so long if they must also attain the technical objectives of the attack.
\vspace{0.2\baselineskip}

None of this comes for free. Large-scale IT projects are famously expensive and slow, and while some of these interventions are relatively cheap others could be extremely complex and time-consuming. Updating business-critical systems can be risky and disrupt core business function, especially when infrastructure is not virtualized or cloud-based, or software has been developed in-house. For code generation and review, defenders will benefit from the independent economic incentives to improve AI's capabilities (namely, the generalized demand for AI-accelerated software engineering). If these advances allow leading-edge companies to more effectively patch vulnerabilities in their own software, then trailing-edge companies who rely on those software offerings will indirectly benefit once those patches roll out.
\vspace{0.2\baselineskip}

Conversely, the integration of inventory and network analysis will be complicated by the fact that each company has a unique technical environment. This makes generalization more difficult, unlike writing and reviewing code (where there are a relatively small number of languages and frameworks). Each company must integrate new defensive capabilities into their bespoke environment, including the associated operational and technical costs. Even once these systems are integrated and functioning, the ongoing cost of using models for continual analysis could be substantial, further slowing the rate of adoption.
\vspace{0.2\baselineskip}

In total, trailing-edge organizations can look forward to falling prices on code generation and review tools, though these tools are not perfect and can do only so much to modernize legacy systems themselves. To integrate these and other tools, companies will need to invest in adoption, whether by altering their software development lifecycle or by allocating staffing to large infrastructure modernization projects. Consequently, we do not believe that this changes the overall perspective for organizations that already struggle to assign sufficient headcount and funding to security, but it does confirm that proactive investment and modernization will benefit from AI's assistance.
\vspace{0.2\baselineskip}

\subsection{Why Be Concerned?}
At this point in our argument, a keen observer might be inclined to ask, ``So what?'' If trailing-edge companies have been lagging behind their peers for years already, accepting the low-but-present risk of cyberattacks, then some of those organizations may continue (intentionally or not) following the same playbook while AI advances. A successful attack undoubtedly has business consequences.\footnote{\citet{huang2023devastating} and \citet{NBER}.} But, the expected cost of such a breach could simply be less than the incremental cost in security funding required to prevent such attacks—if prevention is even possible. Like many commentators have noted, there is a substantial shortage of cybersecurity talent available in many countries,\footnote{\citet{lohnimpact}.} meaning that additional investment may have nowhere to go. Is the rational conclusion for these companies to simply avoid spending until they are certain that adversaries have taken notice of them?\vspace{0.2\baselineskip}

We do not believe trailing-edge organizations should be complacent for three reasons. First, if the rate of attacks ramps up sharply across multiple sectors of the economy, there may be a drastic shortage of cybersecurity talent, limiting the effect of any investment following the attack and prolonging the period of vulnerability. As noted above, many trailing-edge organizations already struggle to staff their security organizations.\footnote{\citet{cisco2021}.} These shortfalls are only expected to grow over time.\footnote{\textit{See} \href{https://www.nist.gov/system/files/documents/2023/06/05/NICE\%20FactSheet_Workforce\%20Demand_Final_20211202.pdf}{NICE fact-sheet}.} With little slack in security hiring, and finite capacity of contractors, there is a real risk of security investment ramping up following a wave of attacks but being unable to hire enough talent regardless. Though improved cybersecurity training programs are undoubtedly necessary, this effect can also be ameliorated by investing early in security practices.
\vspace{0.2\baselineskip}

Second, many of these organizations exist in a state of severe technical and organizational debt. Equifax, as detailed above, spent two years attempting to implement a cybersecurity modernization plan before it was attacked, and had not managed to achieve several key objectives during that time. A single case study does not prove a general trend—but even without mismanaged technical teams, some technical changes can take months or years to implement fully, especially for organizations with sprawling digital operations. Hence, a successful attack does not trigger a sudden reckoning and rapid reversal; it is merely the starting gun for what may amount to years of investment and hiring. This remains true \textit{even if} AI's development yields substantial benefits to defenders, as the process of integrating (for example) AI code reviews, automated code generation, and AI-assisted intrusion detection may take many trailing-edge organizations months or longer. Therefore, if the rate of cyberattacks targeting exposed organizations ramps up quickly, but improving defensive posture remains a long-term project, the total business consequences suffered during that window can be severe.
\vspace{0.2\baselineskip}

Third, there are substantial negative externalities generated by successful attacks. Though data breaches often have direct negative consequences for the targeted business, there are a host of harms which fall directly to consumers or society. For example, leaked passwords can result in account compromise when those passwords have been reused, leaked personally identifiable information (PII) can contribute to identity theft, and leaked credit card details can be resold on the dark web. In some states, consumers can sue for damages, but often there are no consequences besides a bad news cycle and a few firings to demonstrate accountability.\footnote{For example, the California Consumer Privacy Act permits consumers to sue for damages following the disclosure of nonencrypted, nonredacted personal information caused by a failure to maintain reasonable security.} At a societal level, the consequences can be even more substantial: The 2021 Colonial Pipeline hack resulted in panic buying of gasoline throughout the American South,\footnote{\citet{pipelineattack}.} and Britain's National Health Service had their operational practices severely degraded for a day following the WannaCry attack.\footnote{\citet{nhs}.}
\vspace{0.2\baselineskip}

Externalities are, by definition, not priced into an organization's business decisions. Yet, as governments across the world turn a sharper eye towards the risk of cyberattacks, widespread harms to consumers can trigger regulatory action—so even if trailing-edge organizations do not view the existence of negative externalities as a reason to take action on security, they provide additional evidence to support legislative or regulatory action.
\vspace{0.2\baselineskip}

Next, we provide recommendations for both trailing-edge organizations and governments. For organizations, we suggest two reforms to improve the institution's ability to rapidly respond to novel cyber threats: the assignment of authority to an individual responsible for operational cyberdefense decisions; and the active measurement and optimization of patch deployment timelines. We also suggest three reforms to begin identifying and improving vulnerable systems: incorporating vendors' cyber track record in the procurement process; integrating automated security review tools; and creating a comprehensive catalogue of digital assets. For governments, we suggest two complementary policy initiatives. First, to encourage firms to internalize the costs of harms to consumers and societies, we recommend the creation of a private right of action, paired with a statutory floor for ``reasonable'' cyber practices. Second, to help smaller organizations afford assistance, we recommend subsidizing or directly providing cyber audits for these firms.

\subsection{Next Steps for Affected Organizations}
Trailing-edge organizations are faced with substantial work under uncertain timelines and risks. We expect that many of the AI-enabled risks we have detailed are already possible (or will be in the immediate future), given recent research results.\footnote{\textit{See} \citet{cvebench}, \citet{bountybench}, \citet{cybergym}, \citet{cybench}.} The only question is how quickly these technologies will diffuse to motivated attackers, and which organizations will be targeted first. Consequently, time is of the essence—and as noted above, organizations often cannot simply hire more people to address the problem, given substantial shortfalls in security talent. We propose five initial reforms which organizations can take today to begin improving their security posture. These are not comprehensive; to fully assess an organization's vulnerabilities, we recommend organizations pay for comprehensive audits of their security practices and work with domain experts to prioritize and address all technical issues. These reforms are intended as a starting point, and as accelerants for subsequent security-oriented work.

\subsubsection{Assign Organizational Authority}
A predominant priority for trailing-edge organizations is to ensure that security is treated seriously by executives, and make sure that a single individual at the officer or director level has the ability to authorize operational reforms by security staff.\footnote{For an empirical treatment of cybersecurity risk and organizational design in the higher education setting, see \citet{evidence}.} This is, fundamentally, a precursor to most other defensive improvements: As the Equifax case study demonstrates, it is far too easy for large organizations with distributed (or nonexistent) security teams to get bogged down with internal prioritization and turf wars, delaying the delivery of improvements by months or years. There are many ways to assign this authority—but we believe it is important that a single individual possess the responsibility and authority to approve changes to an organization's infrastructure and security practices, and in rare circumstances, override other product priorities when not doing so would lead to substantially increased cyber risk.

\subsubsection{Incorporate Cyber Track Record When Evaluating Vendors}
As discussed above, organizations which rely on third-party software often import the risk profile of the vendor into their own software stack. This places a ceiling on an organization's defensive practices, as once a contract has been signed, there is substantially less leverage over the vendor for any particular security-related decision (not to mention substantially lessened visibility). Organizations should begin by stemming the bleeding: They should ensure that their procurement process incorporates an outside evaluation of the vendor's cybersecurity practices and history of data breaches. Factors to consider include breach history, rapid vulnerability disclosure, and patch frequency.\footnote{In particular, cloud-based solutions (which allows the vendor to apply patches automatically) have categorically faster remediation than on-prem solutions.} These assessments are not bulletproof,\footnote{Vendors are not incentivized to disclose past breaches if doing so may cost them business. As one example of a potential solution, \citet{bair2018close} recommend addressing non-disclosure via the creation of a nationwide reporting system which rewards sharing with safe harbors from government scrutiny. Disclosure of breaches can also be mandated in contractual language, which appears to be growing more common. \textit{See} \citet{acc2025cybersecurity}, 43.} but can, at a minimum, begin orienting vendor decisions toward software companies with a track record of operational excellence. This is a high-leverage change, as vendors with a track record of breaches introduce new sources of risk for purchasers, while vendors with high-quality security practices can replace vulnerable, self-hosted software with well-maintained alternatives.

\subsubsection{Measure Time-to-Deploy for Patches}
An old business proverb reads, ``You can't manage what you can't measure.'' The same goes for many technical objectives, including the time it takes for an organization to deploy security patches to the entirety of their infrastructure. Earlier discussion has demonstrated that delays in patching often provide attackers with the opportunity to infiltrate systems—and as we have argued, AI will enable attackers to develop exploits very quickly following the disclosure of a vulnerability and associated patch, making quick patch timelines critical. Identifying organizational bottlenecks and solving them may require additional headcount or software, but organizations should begin by assessing how much of a problem there is to solve. We recommend organizations define a service-level objective for rollout of critical patches to 100\% of affected machines, then work to converge their practice with this objective. There are tradeoffs for patch deployment speed, of course: Any software update risks incompatibilities with existing systems, and careless rollouts can produce downtime. Organizations should carefully consider which safeguards on deployment are critical, and which are the result of bureaucratic inertia. In any case, though, companies should begin by monitoring before moving to improving patch procedures.

\subsubsection{Begin Integrating AI Review Tools}
Though trailing-edge organizations may not be positioned to reap all the defensive benefits of continuing AI progress, some easy wins are available. For trailing-edge corporations which maintain their own software (as opposed to those who rely solely on third-party tools), AI-assisted code review for the purposes of identifying potential security vulnerabilities is available in many off-the-shelf products and can be easily integrated into existing software development workflows. The highest priorities for review are technical services which are exposed on the open internet, as these services present clear entry points for attackers. This is not a cure-all: AI reviewers can catch only vulnerabilities which can be recognized from the files under review, not those which rely on expansive knowledge of an organization's full codebase. Yet, it is a straightforward, low-budget way to start instilling a security mindset among technical employees and to identify low-hanging vulnerabilities.

\subsubsection{Catalogue Digital Assets}
Finally, organizations must begin by ensuring they know the total surface area of their digital assets. Legacy systems, one-off solutions, and forgotten third-party software can easily be forgotten by employees, as happened with Equifax's legacy Apache Struts deployment. Organizations should establish an authoritative record of all digital assets. Ideally, this should be done via a technical solution—orchestration or centralized monitoring for self-administered infrastructure, and automated asset discovery for third-party software.

\subsection{Next Steps for Government Actors}
Governments cannot afford to let each organization determine whether or not it will invest in cybersecurity, especially when successful attacks often cause harm to citizens. It is difficult to identify one-size-fits-all policy solutions, however, especially when trailing-edge organizations range from globe-spanning credit reporting agencies to small enterprises in the healthcare or education sectors. In all cases, the government should act promptly to change the economics of investing in cybersecurity for each of these organizations. In particular, governments should consider establishing a private right of action for consumers affected by data breaches and subsidizing the provision of cybersecurity services for organizations that otherwise lack the funding and expertise to do so.

\subsubsection{Establish a Private Right of Action for Data Breaches}
Organizations who fail to prioritize cybersecurity generate a host of costs borne by private individuals and governments. If not forced to internalize these costs, it can be a rational decision for these organizations to avoid spending on improving their defensive posture. Governments can incentivize this spending by taking two steps: First, establishing a private right of action allowing individuals to sue companies when their data is lost due to a cyberattack, if that company failed to follow reasonable security practices.\footnote{Not all cyberattacks result in data breaches (e.g. ransomware attacks), but we believe that many organizations which act to secure sensitive data under their control will improve their overall cybersecurity posture in the process. For an argument in favor of expanded liability standards for software, \textit{see} \citet{lawfare}.} Second, establishing statutory standards for what constitutes negligent security practices.\footnote{For a thorough examination of possible ways of defining standards, see \citet{lawfare2}. The California Consumer Privacy Act was recently \href{https://iapp.org/news/a/california-adopts-cybersecurity-audit-rule-outlining-reasonable-cybersecurity}{amended} to require annual cybersecurity audits, which implicitly spells out a floor for what cybersecurity practices constitute the minimum acceptable standards.} These two actions combine to set a floor for how companies must act to secure their internal environments, and allow governments to ratchet up compliance requirements over time.\footnote{It is always possible that statutory standards will not be updated over time, so these standards should be treated as a floor and not a ceiling for compliance purposes.}  This policy is particularly impactful for larger organizations, who concern themselves more with litigation risk (and likely employ lawyers who are responsible for assessing that risk); the potential liability for smaller organizations may not be sufficient to drive change, or they may simply not be aware of the risk of noncompliance.

\subsubsection{Subsidize Security Services for Enterprises Missing In-House Expertise}
Many small- and medium-sized businesses lack a dedicated security function, and instead have only general internal technology roles or similar. Consequently, they may struggle to prioritize security while ``keeping the lights on'' for the rest of their business operations. Governments should consider subsidizing or providing security services to such enterprises; audits could be done both by the government itself (where the capacity exists) or by private actors, with the cost partially borne by the government itself. These subsidies match the classic responsibilities of government: To provide public goods (namely, reliable and secure digital services) that the market otherwise fails to account for. This reform can be paired with policies encouraging development of additional cyber talent to meet the substantial need, such as via tuition credits or tax writeoffs for workplace development costs.

\subsubsection{Invest in Technologies to Improve Defensive Postures}
The government might also consider investing in the development of new defensive technologies that either help trailing-edge organizations benefit without substantial independent investment, or which allow third-party vendors to accelerate the speed of patching and remediation for their customers.\footnote{For examples, see DARPA's recent \href{https://www.darpa.mil/research/programs/enhanced-sbom-for-optimized-software-sustainment}{announcement} of research into scaled vulnerability detection and remediation, as well as \href{https://www.darpa.mil/research/programs/compartmentalization-and-privilege-management}{automated partitioning} of software systems to reduce the privileges granted to actors. The U.S. federal government also provides direct funding for state and local governments to address cybersecurity risks, though not to private entities.} These technologies offer the potential to improve the defensive posture of all enterprises without requiring punitive actions, and without requiring proactive buy-in from each vulnerable firm. Notably, though we are focused on government action for the purposes of this section, many nonprofit and academic researchers are similarly focused on identifying highly-scalable uses of AI to improve the cyber posture of all of society\footnote{\textit{See, e.g.}, \citet{bradleyrefactor}, \cite{councilman2025formalverificationllmgeneratedcode}.}—and governments can indirectly support this work via grant-making and other institutional support.

\section{Conclusion}
AI's continued advancement promises to substantially impact both offensive and defensive capabilities for cyber. A common frame for analyzing these effects is via the ``offense-defense balance,'' which examines the comparative cost to attackers versus defenders to attain victory. This framing relies on defenders being willing to invest proportionally to the risk of attack, and hence, proportionally to attackers' investment. Yet, many organizations today invest only minimally in cybersecurity, plausibly because they do not believe attackers are likely to target them, or alternatively because organizational gridlock inhibits their ability to improve their defensive posture on a relevant timescale. These enterprises, which we call ``trailing-edge organizations,'' do not effectively internalize the cyber risk from attackers—and hence, the discussion of whether AI's advancement will privilege attackers or defenders is largely immaterial to their cyber practices.
\vspace{0.2\baselineskip}

Many trailing-edge organizations have escaped notice from attackers for three reasons: The number of attacks which may be launched is limited by human bandwidth to identify targets, develop exploits, and process exfiltrated data; the number of humans with the requisite knowledge (and willingness) to launch an attack is quite limited; and some would-be attackers are deterred by the substantial risk of real-world consequences enforced by law enforcement. Each of these assumptions, however, is primed to change in the near future. The widespread availability of capable AI systems, and especially those that can be misaligned or jailbroken, means that adversaries will be able to minimize the role of humans in the cyber kill chain; that more individuals will be able to acquire the knowledge needed to launch an attack; and that law enforcement will struggle to identify the perpetrators of cyberattacks.
\vspace{0.2\baselineskip}

AI's capabilities also apply to improved technical capacities for attackers. In particular, attackers will be able to translate vulnerability disclosures into working exploits more rapidly than is possible today, meaning that any delay in applying security patches comes with heightened risk for defenders. For open-source software, attackers will be able to observe the vulnerability patching process and potentially develop an exploit before a patch is widely available to users. It will also become easier for both attackers and defenders to rapidly scan an application for similar vulnerabilities once a single vulnerability is disclosed—but for trailing-edge organizations (or, similarly, under-resourced open-source software maintainers), attackers will retain a substantial edge. Finally, target identification will benefit from AI. Following the development of an exploit, it will be easier to identify all vulnerable organizations, tailor the exploit as needed for each, and launch multiple attacks in parallel.
\vspace{0.2\baselineskip}

The lowered cost of launching the marginal cyberattack intersects with this heightened technical ability for attackers to produce a uniquely dangerous world for trailing-edge organizations. These organizations already struggle to adhere to cyber best practices, and in many cases, may need to invest months or years of effort to modernize their systems. Their defensive capabilities may benefit from AI, but human-scale failures such as organizational inertia, legacy code, and overreliance on humans will limit those benefits—and correspondingly increase the importance of defensive investment in the immediate future, before attacks become vastly more common.
\vspace{0.2\baselineskip}

By default, many trailing-edge organizations only increase their investment in security after falling prey to a cyberattack. This approach is unlikely to succeed in a near-term future featuring widely-deployed advanced AI systems, as these investments will likely be hamstrung by talent bottlenecks, take years to come to fruition, and generate substantial negative externalities in the interim. It is important for organizations to act now to modernize their cyberdefense practices and position themselves to be able to realize continued defensive uplift from improved AI capabilities. Organizations should take low-cost steps today, such as ensuring a single executive or director is accountable for cybersecurity, evaluating vendors on the basis of their cyber track record, and measuring their current time-to-deploy for security patches. Governments can encourage this transition by lowering the cost of improved cyberdefense or by making these organizations internalize the societal costs generated by poor defensive practices.
\vspace{0.2\baselineskip}

The net effect of AI for attackers versus defenders is not yet clear for leading-edge organizations, who are willing to invest in defense proportionally to the risk posed by attackers. Focusing on this offense-defense balance, however, obscures the glaring deficiencies in cyberdefense present for most organizations, and the need for immediate investment to bring those organizations into compliance with existing cyber best practices. Helping these organizations solve these weaknesses cannot be left to a wait-and-see approach—the potential harms are spread across society, and demand immediate action by both affected companies and government actors.

\section{Acknowledgments}
This work was done while both authors were Summer Fellows at the Centre for the Governance of AI, and was presented at the Technical AI Governance Forum in August 2025. We thank Jeffrey Ding, Matthew van der Merwe, John Halstead, Alan Chan, Herbie Bradley, Elias Groll, and Catherine Chen for helpful comments and discussion. All errors are our own.

\bibliographystyle{plainnat}
\bibliography{references}

\end{document}